\documentclass[sn-aps]{sn-jnl}% Basic Springer Nature Reference Style/Chemistry Reference Style
\usepackage{graphicx}%
\usepackage{filecontents}
\usepackage{multirow}%
\usepackage{amsmath,amssymb,amsfonts}%
\usepackage{amsthm}%
\usepackage{mathrsfs}%
\usepackage[title]{appendix}%
\usepackage{xcolor}%
\usepackage{textcomp}%
\usepackage{manyfoot}%
\usepackage{booktabs, tabularx}%
\usepackage{xr}%
\usepackage{algorithm}%
\usepackage{algorithmicx}%
\usepackage{algpseudocode}%
\usepackage{lmodern}
\usepackage{listings}%
\usepackage{float}
\makeatletter

\newcommand*{\addFileDependency}[1]{% argument=file name and extension
\typeout{(#1)}% latexmk will find this if $recorder=0
\@addtofilelist{#1}
\IfFileExists{#1}{}{\typeout{No file #1.}}
}\makeatother

\newcommand*{\myexternaldocument}[1]{%
\externaldocument{#1}%
\addFileDependency{#1.tex}%
\addFileDependency{#1.aux}%
}

\myexternaldocument{Supplementary}
\begin{document}

\title[Article Title]{Simulating metal complex formation and dynamics in aqueous solutions: Insights into stability, mechanism, and rates of ligand exchange}


\author[1,2,3]{\fnm{Luca} \sur{Sagresti}}%\email{iauthor@gmail.com}
%Uncomment line below for equal contribution

\author[1,2]{\fnm{Luca} \sur{Benedetti}}
%Uncomment line below for equal contribution

\author[4,5]{\fnm{Kenneth M.} \sur{Merz Jr.}}
%Uncomment line below for equal contribution

\author*[1,2,3]{\fnm{Giuseppe} \sur{Brancato}}\email{giuseppe.brancato@sns.it}
%Uncomment line below for equal contribution

\affil[1]{\orgname{Scuola Normale Superiore}, \orgaddress{\street{Piazza Dei Cavalieri 7}, \city{Pisa}, \postcode{I-56126}, \country{Italy}}}
\affil[2]{\orgname{Istituto Nazionale di Fisica Nucleare (INFN)}, \orgaddress{\street{Largo Pontecorvo 3}, \city{Pisa}, \postcode{I-56127}, \country{Italy}}}

\affil[3]{\orgname{Consorzio Interuniversitario per Lo Sviluppo Dei Sistemi a Grande Interfase (CSGI)}, \orgaddress{\street{Via Della Lastruccia 3}, \city{Sesto Fiorentino (Fi)}, \postcode{I-50019}, \country{Italy}}}

\affil[4]{\orgdiv{Department of Chemistry}, \orgname{Michigan State University}, \orgaddress{\street{Street}, \city{East Lansing}, \postcode{48824}, \state{Michigan}, \country{United States}}}

\affil[5]{\orgdiv{Department of Biochemistry and Molecular Biology}, \orgname{Michigan State University}, \orgaddress{\street{Street}, \city{East Lansing}, \postcode{48824}, \state{Michigan}, \country{United States}}}


\abstract{Metal coordination is ubiquitous in Nature and central in many applications ranging from nanotechnology to catalysis and environmental chemistry. Complex formation results from the subtle interplay between different thermodynamic, kinetic, and mechanistic contributions, which remain largely elusive to standard experimental methodologies and challenging for typical modeling approaches. Here, we present an effective molecular simulation approach that can fully describe the chemical equilibrium and dynamics of metal complexes in solution, with atomistic detail. Application to Cd(II) and Ni(II) complexes with various amine ligands provides an excellent agreement with available association constants and formation rates spanning several orders of magnitude. Moreover, investigation of polydentate ligands allows unravelling the origin of the chelate effect as due to the concurrent contribution of entropy, dissociation rates, and ligand binding mechanisms. This study represents a step forward for the {\it in silico} design of coordination chemistry applications and for a better understanding of biochemical processes activated by metal binding.}




\maketitle
%%%%%%%%%%%%%%%%%%%%%%%%%%
%Comment this line to remove line numbering
%\linenumbers
%%%%%%%%%%%%%%%%%%%%%%%%%%

%\textbf{Requirements for Nat Chem:} 3k words for intro, results and conclusion. no more than 3k for methods. max 6 items in the main and 10 in the supp. \\

\section*{Main}\label{sec1}  % WORD COUNT: 515
%many references needed but put after the intro has been more or less outlined

Metal complexes play pivotal roles across diverse domains of chemistry, encompassing catalysis \cite{herrmann1993,hammer2017}, biochemistry \cite{handel1993,Joseph2014}, materials science \cite{FRIESE2008,wang2020b}, and applications in analytical and environmental chemistry.\cite{CIESLA2004}
In recent studies, metal coordination was exploited to design chelating polymers targeting specific transition metals.\cite{diallo2008,Fortuna2015,GUO2019,hruby2021,GHISALBERTI2020} In molecular biology, metal complexes have been used to probe nucleic acid structures\cite{pyle1990}, to enable site-specific cleavage,\cite{YU201837, RODRIGUEZ2021} and to develop anticancer drugs.\cite{Frezza2010, ndagi2017, Hu2022} In nanotechnology, they have been investigated for their potential application as structural and electron-transfer probes.\cite{aron2015, BERRONESREYES2021} \\

Recent literature has also clearly highlighted the need for a deeper molecular-level understanding of metal coordination. Tuning the enthalpy-entropy balance dictating complex stability is, for example, crucial to improve the catalytic activity and selectivity of organometallic catalysts.\citep{moschetta2014} At the same time, unraveling the subtle mechanisms by which metal complexes can activate or hinder fundamental molecular processes allows us to obtain significant breakthroughs in various fields, such as biology\citep{liao_structural_2012, spatzal_ligand_2014, dudev_competition_2014} and environmental chemistry \citep{ding_development_2017, ravi_origin_2017, gao_adsorption_2020}. Finally, the ability to investigate the detailed thermodynamic and mechanistic factors underpinning metal complex formation and stability is essential when fostering advances in coordination chemistry. \\

In this context, {\it in silico} studies are, in principle, well-suited for gaining atomistic insights into metal binding complexes, especially for structural details.\cite{Deeth2009, Pengfei2017, Nandy2021} For example, metal ion-ligand interactions can be properly modeled by electronic structure methods\cite{vallet2003, uudsemaa2003} or by carefully optimized force fields\cite{Sengupta2018}. However, the realistic simulation of a metal complex's equilibrium (i.e., M $\rightleftharpoons$ ML  $\rightleftharpoons$ ML$_2$ $\rightleftharpoons$ ...) and ligand exchange dynamics has proven to be far more challenging\cite{hyla-kryspin2004, deyonker2006, jiang2012, weaver2013} owing to the long timescale of complex formation and ligand exchange, which are typically $> 10^{-6}$ s.\cite{martell1996, smith1989} 
Here, a purposely developed {\it in silico} approach was used to provide, for the first time, a complete picture of the chemical equilibrium and kinetics of metal complexes in solution. The approach combines a recently finely-tuned interatomic potential developed by Sengupta et al.\citep{Sengupta2018}, rooted in the 12-6-4 non-bonded Lennard-Jones potential\citep{Li2014, Li-Song2014} which proved effective in modeling a range of transition metal assemblies\citep{li2023, jafari2024}, with state-of-the-art enhanced sampling (i.e., metadynamics and its variants\citep{laio2002, barducci2011, Bussi2020}) and kinetics (i.e., Markov State Model\citep{PANDE2010, prinz2011, husic2018}) techniques. As a test case, a series of cadmium (Cd$^{2+}$) and nickel (Ni$^{2+}$) complexes in water with amine ligands of variable denticity and length were considered. The thermodynamic equilibrium between all the stable and metastable ternary species ML$_n$S$_m$ (M: metal ion, L: ligand, and S: solvent, with $n,m$=0,1,...N) was obtained, showing the relative stability, the energy barriers for interchange, and the minimum free energy pathways leading to complex formation. In addition, mechanistic insights not easily accessible by experiments were also provided, for example, unraveling the nature (i.e., associative or dissociative) of the ligand substitution reaction and the role of the solvent. Both estimated stability constants (i.e., $K_i$ = [ML$_i$]/[ML$_{i-1}$][L]) and association/dissociation rates reproduced very well their experimental counterparts, spanning several orders of magnitude, with great accuracy. Importantly, the results shed more light on the interplay between entropy, ligand unbinding rates, and mechanistic contributions underpinning the chelate effect of polydentate amine ligands. \\

\section*{Results}\label{sec2}   % WORD COUNT: 3245

\subsection*{Simulation of metal complex equilibrium in solution}\label{2A}
%% Fig.1A 2D map
%% Fig.1B DG wrt free M and intermediates
%% Fig.1C pKa
%% Fig.1D DG all conc. and ratios
%% SI \textbf{Table S\ref{tab:Fs_diffConc}} 
%% SI \textbf{Table S\ref{tab:pKs_diffConc}}

An aqueous solution of Cd$^{2+}$ (0.05M) and ethylenediamine (0.15M, hereafter referred to as en) at normal conditions was adopted as a prototypical model of metal coordination to illustrate the proposed computational approach. Applications to other amine ligands and/or concentrations are reported on below. In analogy with experiments, we ignored the presence of protonated amine species, assuming a moderately basic solution, and we neglected the counterion's participation in complex formation (see Methods).
Metal coordination complexes were described in terms of both water and amino coordination numbers, ML$_n$S$_m$ (i.e., for polydentate amines, each amino moiety was counted separately). By performing extended metadynamics simulations, the 2D free energy landscape of the equilibrium solution of the Cd(II)-en complex was obtained (Fig. \ref{fig:fig1}a). The map illustrates the relative stability of all chemical species formed under the given physico-chemical conditions and further unravels valuable information about the system, such as the minimum free-energy pathway to complex formation, the interchanging energy barriers, and the nature of the ligand-solvent exchange mechanism (whether associative or dissociative) between different coordination species. Fig. \ref{fig:fig1}c shows that [Cd(en)$_2]^{2+}$ and [Cd(en)$_3]^{2+}$ complexes are the most favorable ones ($\approx$ -30 kJ/mol) with respect to the free metal Cd(II), as compared to the monocoordinated species, [Cd-en]$^{2+}$ (-17.2 kJ/mol). Fig. \ref{fig:fig1}c also shows the interesting free energy pattern as a function of NH$_2$ coordination, highlighting the relative higher stability of even coordination numbers (i.e., n(NH$_2$) = 2, 4, 6) with respect to odd ones (i.e., n(NH$_2$) = 1, 3, 5), a result highlighting the bidentate ligand's preference for chelating ring configurations (i.e., five-membered ring structures).

For comparison with experiments, we evaluated the thermodynamic association constants (i.e., $pK_i$, with $i = 1, 2, 3$) from the complex equilibrium population, obtaining a remarkable agreement with experiment(Fig.\ref{fig:fig1}e). The computational methodology was further assessed by estimating the relative $\Delta G$(ML$_i$ - ML$_{i-1}$) from the experimental $pK_i$ at various Cd(II)-en concentrations and ratios. Data are reported in Supplementary Table \ref{tab:pKs_diffConc} and \ref{tab:Fs_diffConc}. Overall, the simulation results showed an excellent agreement with the experimental data (i.e., mean absolute error $<$1 kcal/mol), with some noticeable deviations observed only in the case of the 1:1 and 1:2 metal-ligand ratio system, for which sampling the ML$_3$ population was particularly challenging (Supplementary Fig. \ref{fig:multi-Cd-en}). Hence, the present approach is robust, provided that the ligand is in excess relative to the metal ion (at least three times larger), though this is also usually the case in the experimental investigation.
Note that the equilibrium properties discussed above reflected the remarkable accuracy of the underlying Cd(II)-en interaction model (a result not necessarily expected since the original force field was optimized towards the interaction of a metal ion with one ligand only\citep{Sengupta2018}). Yet, water coordination was somewhat overemphasized by the model (8-fold vs the usual 6-fold coordination), a result unrelated to the proposed simulation approach and with minor impact on the resulting thermodynamic analysis.
In analogy with Cd(II), we also investigated the Ni(II)-en complex which is known to be characterized by a higher stability (x100) than Cd(II)-en. Again, we obtained similar free energy profiles and a favorable agreement with experimental association constants (Figs. \ref{fig:fig1}b, \ref{fig:fig1}d, and \ref{fig:fig1}f). It is worth noting that despite the strong stability of the [Ni(en)$_3]^{2+}$ species, both enantiomers (i.e., the left-handed and right-handed propeller illustrated in Fig. \ref{fig:fig1}g) were observed to form and disassemble in our simulations.  

\subsection*{Ligand effects on complex stability}\label{2B}
%% RESULTS
%% Fig.2A,B,C 2D maps: nme, dien, put 
%% Fig. 2D pKa: histogram nme
%% Fig. 2E pKa: histogram dien, put
%% Table 1 all thermo data

%% Fig.2F DG wrt free M and intermediates
%% Fig.1D DG all conc. and ratios
%% SI \textbf{Table S\ref{tab:Fs_diffConc}} 
%% SI \textbf{Table S\ref{tab:pKs_diffConc}}

The same computational procedure outlined above was extended to investigate methylamine (nme) and diethylenetriamine (dien), to obtain a progressive series of mono-, bi-, and tridentate ligands,  and putrescine (put, 1,4-diaminobutane) a bidentate compound with a longer chain than en. To make a fair comparison between all systems, we considered the same ratio (1:6) between metal ions and ligand amino groups (i.e., 1:6 for Cd(II)-nme, 1:3 for Cd(II)-en/put, and 1:2 for Cd(II)-dien).
The computed free energy maps (Fig.\ref{fig:fig2}a, b, and c) showed an enhanced preference for ligand coordination with increased denticity: at equilibrium, the most favorable complex configurations are those with 2 and 3 bonded NH$_2$ for nme, 4 and 6 for en, and 6 for dien. 
$\Delta$G(ML$_n$) follows a regular pattern for nme and put, in contrast to en and dien where only configurations displaying fully-coordinated ligands are stable at equilibrium (Fig.\ref{fig:fig2}d-f).
As seen for en, Cd(II)-dien configurations displaying at least one free (not bonded) NH$_2$ group were rather unstable. Cd(II)-put showed an intermediate behavior between nme and en (Fig.\ref{fig:fig2}c and f), since the increased chain length and flexibility make this ligand more similar to the monodentate situation. Cd(II)(put)$_2$ was the most stable complex in solution (-17.2 kJ/mol), showing comparable stability with the tetra-coordinated Cd(II)(nme)$_4$ (-14.0 kJ/mol) but rather different than Cd(II)(en)$_2$ (-30.5 kJ/mol). \\

$pK_i$'s issuing from simulations and experiments displayed a remarkable agreement (Fig.\ref{fig:fig2}g-i). For nme, p$K_i$'s were all very well reproduced, with a slight deviation for pK$_4$ (about 0.5). Some discrepancies were obtained for put and dien since in these models the metal-ligand interaction potential was borrowed from nme and en, respectively, without further reoptimization (Methods). However, the stability trend provided by the association constants along the amine series was well reproduced: for p$K_1$, nme $<$ put $<$ en $<$ dien (also, p$K_1$(en) $>$ p$\beta_2$(nme) $>$ p$K_1$(put) and p$K_1$(dien) $>$ p$\beta_3$(nme)). In particular, the chelate effect of en and dien, as defined in the seminal paper by Schwarzenbach\citep{Schwarzenbach1952}, was in excellent agreement with experiments: p$K_1$(en) - p$\beta_2$(nme) = 0.6 (exp: 0.6); p$K_2$(en) - p$\beta_4$(nme) + p$\beta_2$(nme) = 2.73 (exp: 2.79); p$K_1$(dien) - p$\beta_3$(nme) = 1.32 (exp: 1.96). \\

In Table \ref{tab:thermo-table}, we report the enthalpic ($\Delta$H) and entropic (-T$\Delta$S) terms of the most relevant chemical species throughout the amine series, allowing a more direct comparison between different metal complexes. When normalized for the number of amino groups, all ligands examined showed a similar $\Delta$H contribution ($\approx$ -21 kJ/mol), with the exception of Cd(II)-dien ($\approx$ -17.6 kJ/mol),  for the reasons noted above. 
In contrast, significant differences were observed in the (normalized) entropic contribution: a steady decrease versus ligand denticity was observed in going from nme to dien (i.e., on average 10.9, 8.4 and 5.9 kJ/mol for nme, en, and dien, respectively). T$\Delta\Delta$S for going from Cd(II)-(nme)$_2$ to Cd(II)-en and from Cd(II)(nme)$_4$ to Cd(II)(en)$_2$ were about 18 and 45 J/mol, respectively, in line with experiments and larger than the corresponding enthalpic differences (Fig. \ref{fig:fig2}j,k). Overall, these results are consistent with previous experiments\citep{Spike1953,smith1989,martell1996} and illustrate the entropic origin of the chelate effect of bi- and tridentate amine ligands. As expected, the enhanced rotational flexibility of put resulted in a larger entropic term than for the other ligands (13.8 kJ/mol).

\subsection*{Water-ligand exchange mechanisms}\label{2C}

% Dynamical analysis
% Water Coord. Num. (font larger) anche numeri [1.0  0.0]???
% Amino Coord. Num.
% Leaving Water Prob. Dist.
% Add a circle for coordination shell boundary
% 1st binding (larger font)
% 2nd binding (larger font)
% Ang instead of nm
% Prob. Dist scale, why 0 to 16?
Metal complexes in aqueous solutions are formed through a substitution reaction between water and the ligand in the first ion coordination shell,\citep{Huheey2006} where according to Langford and Stengle's classification\citep{Langford1968} such an exchange follows either an associative or a dissociative mechanism.
Previous experimental studies on polyamine complexes with transition metal ions supported a dissociative mechanism, starting from the pioneering work of Eigen and coworkers.\citep{Eigen1961,Eigen1965} In particular, kinetic studies hinted that the loss of a water molecule from the first coordination shell represents the rate-determining step in metal complex formation,\citep{Eigen1965,Rorabacher1966,Rorabacher1971,Roche1974} based on the close agreement between the water exchange and the first-ligand binding rates.\citep{Eigen1965,Lincoln1995} \\

For all Cd(II) systems under scrutiny, we observed that the release of a coordinating solvent molecule into the bulk solution preceded the first ligand binding event, as shown by the minimum free energy pathway of complex formation (Fig.\ref{fig:fig1}a and Fig.\ref{fig:fig2}a-c), though the fine details were different, as discussed in the following. 
We analyze in some detail the binding mechanism of ethylenediamine to a free Cd$^{+2}$ from multiple unbiased simulations by closely following the coordination shell around the metal ion for a few picoseconds before and after the binding event (Methods). The coordination of the bidentate ligand occurred as a two-step reaction: first, we observed the binding of one NH$_2$ group to the metal ion (Fig. \ref{fig:fig3}a), then the second amino moiety entered within the first coordination shell of Cd$^{+2}$ (Fig. \ref{fig:fig3}a). Monitoring the evolution of the coordinating water molecules around the metal center, we observed a dissociative mechanism upon first NH$_2$ binding, with a water molecule leaving the Cd$^{+2}$ coordination shell before the ligand (Fig. \ref{fig:fig3}a). However, the second amino binding event, leading to a fully coordinated bidentate ligand, followed either an associative or a dissociative mechanism (Fig. \ref{fig:fig3}b). The associative mechanism led to a temporary over-coordination of Cd$^{+2}$ during the en chelating ring formation. Similar considerations apply when a second en molecule binds to Cd$^{+2}$ (Supplementary Fig. \ref{fig:mechanism_en}). On the otherhand, Figure \ref{fig:fig3}c shows that only a dissociative mechanism is feasible upon the second amino binding event when considering nme as a ligand (i.e., Cd(II)-nme $\rightarrow$ Cd(II)(nme)$_2$), thus highlighting another difference between polydentate ligands versus monodentate ones.

Furthermore, we analyzed the relative orientation of the leaving water molecule with respect to Cd$^{+2}$ and the entering (NH$_2$) coordination group by projecting the coordinates of Cd$^{+2}$, H$_2$O and NH$_2$ on the XY plane, while keeping the metal ion at the origin and the amino group along the X-axis. The analysis focused on the distribution of the leaving water molecule during the first and second binding event of Cd(II)-en formation and the second binding event of Cd(II)(nme)$_2$. In all events, the leaving H$_2$O left the coordination shell via a direction at about $90^\circ$ with respect to the entering amino moiety (Fig. \ref{fig:fig3}d-g). The only observed difference was that in the Cd(II)-en associative mechanism the water molecule resides inside the first coordination shell when NH$_2$ binding occurred (Fig. \ref{fig:fig3}f).

\subsection*{Kinetic analysis of ligand binding and unbinding}\label{2D}

The kinetic analysis of complex formation and ligand exchange was carried out by using Markov State Models (MSMs), which allows the estimate of rate constants spanning several orders of magnitude not readily accessible by standard MD simulations. First, among the several microstates identified by MSM, the PCCA+\citep{Roblitz2013} analysis identified a few coarse-grained clusters that match the complex coordination states (i.e., ML$_n$) previously obtained from the thermodynamic analysis (Fig.\ref{fig:fig4}a,b). Then, the mean first passage times among the different states were evaluated along with the corresponding rate constants (Methods). Fig. \ref{fig:fig4}c-e report the computed rate constants for all complex association/dissociation reactions concerning Ni(II)-en, Ni(II)-nme and Cd(II)-en.

We observed that Ni(II)-en complex formation (i.e., forward) rates were consistently greater than the corresponding dissociation (i.e., backward) rates, with a ratio of k$_{i}$ / k$_{-i}$ spanning a range from 10$^{3}$ to 10$^{7}$. The backward reactions were particularly slow in agreement with the stability of Ni(II)-en complex species. Remarkably, experimental measurements obtained from stopped-flow techniques on Ni(II)-en complex solutions reported data in very good agreement with the present predictions. We obtained k$_{1}$ = 2.8 $\cdot$ 10$^{6}$ M$^{-1}$ s$^{-1}$ and k$_{-1}$ = 0.26 s$^{-1}$ for the first ligand binding and unbinding event, as compared to the experimental rates k$_{1}$ = 3.5 $\cdot$ 10$^5$ M$^{-1}$ s$^{-1}$ and k$_{-1}$ = 0.08 s$^{-1}$).\citep{Taylor1974} Also, an excellent agreement was found for the third ligand association/dissociation rates, where we found k$_{3}$ = 1.0 $\cdot$ 10$^{4}$ M$^{-1}$ s$^{-1}$ (exp: 1.1 $\cdot$ 10$^{4}$ M$^{-1}$ s$^{-1}$) and k$_{-3}$ = 31 s$^{-1}$ (exp: 38 s$^{-1}$\citep{Jones1970}). Besides, it is worth noting that both experiments and simulations displayed similar forward reaction (k$_{1}$) and water exchange rates (k$^{H_2O}$ = 1.8 $\cdot$ 10$^5$ s$^{-1}$, exp: 3.37 $\cdot$ 10$^4$ s$^{-1}$\citep{Ducommun1979}), supporting the view that the release of a water molecule from the first coordination shell is the rate-determining step for complex formation (i.e., dissociation mechanism).\citep{Eigen1965,Richens2005} This was further corroborated by results obtained for Cd(II)-en (k$_{1}$ = 1.4 $\cdot$ 10$^{10}$ M$^{-1}$ s$^{-1}$ vs k$^{H_2O}$ $=$ 5 $\cdot$ 10$^{10}$ M$^{-1}$ s$^{-1}$), though showing a much faster kinetics (about x10$^{4}$) than Ni(II) (Fig. \ref{fig:fig4}e). 

As a further manifestation of the chelate effect, the kinetic analysis highlighted that the backward reactions ML$_i$ $\rightarrow$ ML$_{i-1}$ for the monodentate ligand were about five orders of magnitude greater than the corresponding one for en, while the forward reaction rates were about the same (Fig. \ref{fig:fig4}c,d, Supplementary Table \ref{tab:rates_all}). This finding corroborates the notion that the enhanced stability of metal complexes with chelating agents is due to slower dissociation rates.\citep{Richens2005} Moreover, the relatively low dissociation rates of the bidentate ligand appeared in qualitative agreement with the higher energy barrier experienced in going towards odd coordination number configurations, thus making the disassembly of the Ni(II)-en or Cd(II)-en complexes kinetically unfavorable. Note that stability constants (pK$_i$) evaluated from the k$_i$/k$_{-i}$ ratio aligned fairly well with those obtained from the previous thermodynamic analysis (Supplementary Table \ref{tab:pKs_Cden_from_rates}) and similar rate constants were obtained when testing different metal ion-ligand concentrations (Supplementary Table \ref{tab:rates_Cden_diff_conc}), as further proof of the robustness of our model.

\section*{Conclusions}\label{sec3} % WORD COUNT: 181

In this work, we reported a comprehensive molecular view of metal complex equilibrium in water obtained using a simulation approach based on state-of-the-art enhanced sampling and MSM methodologies. This approach makes the detailed thermodynamic and kinetic analysis of metal coordination systems computationally feasible, thus allowing direct comparison with experiments. Results obtained for a series of metal amine complexes showed an agreement with observed stability constants (p$K_i$) and relative free energies ($\Delta G$(ML$_i$)) within chemical accuracy (1 kcal/mol). Complex formation and ligand exchange rates reproduced available experimental results, thus supporting application to a large number of systems for which kinetic information is presently lacking. Noteworthy, simulations of Cd(II)-amine complexes with variable ligand denticities allowed us to fully underscore the nature of the chelate effect as characterized by the interplay between entropic contributions, dissociation rates, and ligand binding mechanisms. The ability to gain valuable insights into complex formation and ligand exchange makes this computational approach well-suited for aiding metal coordination design in nanotechnological applications and for better investigating metal binding in various biological processes.

\section*{Methods}\label{sec4}

%IN figure: Conversely, when the concentration of the free ligand is very low,  minor errors in the evaluation of $\Delta G$ are amplified in the estimate of the association constants ($pK_i$), as clearly shown by eq. \ref{eq:stab_constant}.
%dien underestimated by our model w.r.t. experiments\citep{smith1989},  

\subsection*{Molecular dynamics simulations}\label{sec:methods_MD}

Various aqueous solutions of Cd$^{2+}$ (0.05M) and ethylenediamine (en: 0.05M, 0.1 M, 0.15 M, 0.225 M), methylamine (nme: 0.16 M, 0.30 M), diethylenetriamine (dien: 0.10 M) and putrescine (put: 0.15 M) were considered. To further assess the size consistency of the thermodynamic analysis, two additional Cd(II)-en systems were prepared with same ratio (1:3), one twice the other. A solution of Ni(II)-en (en: 0.08 M, ratio 1:3) and Ni(II)-nme (en: 0.16 M, ratio 1:6) was also prepared to probe the methodology against a system that experimentally has shown slower exchange times. The complete list of systems is shown in Supplementary Table \ref{tab:sim-recap}.
The 12-6-4 Lennard-Jones type model developed by Li and Merz\citep{Li2014} was adopted for treating the metal-ligand and metal-water interactions since it was successfully tuned to properly reproduce a range of experimental observables, such as structural properties, hydration free energies, and binding affinities.\citep{Li-Song2014} In particular, the Cd(II)-en and Ni(II)-en interaction model taken from ref.\citep{Sengupta2018} was employed in this study and the same metal-nitrogen interaction retained for the dien ligand. On the contrary, the polarizability parameter of the model\citep{Sengupta2018} was slightly increased (from 3.16 to 3.35) in the case of Cd(II)-nme (and Cd(II)-put), to favor the formation of complexes with high coordination numbers (i.e., Cd(II)(nme)$_4$) as observed experimentally with an excess of monodentate ligand\citep{Spike1953}. 
Note that the model overestimates water coordination (8-fold vs 6-fold\citep{marcus1988}) around Cd(II), even though it reproduced accurately the solvation free energy and ion-water distance\citep{Li2014} and it was fruitfully applied in previous studies.\citep{panteva2015,li2023,jafari2024,Zhang2024}
The TIP3P\citep{MacKerell1998} water model was used for modeling the solvent and chloride ions were added to ensure electric charge neutrality of the systems under consideration. Since, in experiments, ClO$_4^-$ and NO$_3^-$ are employed to prevent any significant interaction with the metal and the formation of spurious complexes, we adopted a customized Cd-Cl 12-6 LJ nonbonded model to avoid the formation of ionic couples. Note that experimentally\citep{Paoletti1984} a stable ionic medium is utilized to keep the activity coefficients of a particular ion constant by introducing a high concentration of a specific anion (e.g., ClO$_4^-$ or NO$_3^-$), which is intended to be unreactive and not form any complexes with the ions under study. By doing so, the activity coefficients of the ions being studied can be considered to remain constant in all the solutions \citep{Tobias1958}.\\
All simulations were performed with Amber22\citep{case2022} enforcing periodic boundary conditions and using the PME\citep{essmann1995} algorithm to treat long-range interactions with a 12 \AA\,cutoff (10 \AA\,for the smaller systems).

Minimization was performed using 20000 steps of steepest descent followed by 10000 steps of conjugate gradient. A 250 ps NPT heating procedure was performed to heat the system from 0 K to 300 K followed by a 1 ns equilibration at 300K with constant $NPT$ conditions setting the pressure at 1 atm using Berendsen barostat.\citep{Berendsen1984} The equilibrated geometries were used for the production runs of 100 ns each. Last frame geometries and coordinates were used as a starting point for the metadynamics simulations (see next paragraph). An integration time step of 1 fs was used in the heating step and 2 fs in the production runs. Langevin dynamics temperature\citep{Loncharic1992} control was employed in the heating and the production runs with a collision rate equal to 1.0 ps. SHAKE algorithm\citep{RYCKAERT1977} was applied to constrain the covalent bonds with hydrogen atoms in the amine ligands in all simulations. 

\subsection*{Thermodynamic analysis}

The stability (or association) constants of metal complexes were obtained directly from the equilibrium populations of the corresponding chemical species, through 
\begin{equation}\label{eq:eq_const}
K_i = \frac{[ML_i]}{[ML_{i-1}][L]}    
\end{equation}

and, as usual, reported in terms of their logarithm (i.e., p$K_i = log K_i$).\citep{Paoletti1984}\\
Metal coordination was quantitatively evaluated through a simple and physically sound collective variable originally proposed in ref. \citep{brancato2011,Sagresti2022} proved effective for studying the water coordination around aqua ions. Accordingly, both water and ligand coordination numbers have been considered, the former based on the Cd$^{+2}$-oxygen (water) distance with a cutoff radius of 3.2 \AA\hspace{0.01em} (2.9 \AA\hspace{0.01em} for Ni$^{+2}$), while the latter based on the Cd$^{+2}$-nitrogen (ligand) distance (cutoff radius of 3.4 \AA\hspace{0.01em}).
The well-tempered metadynamics\citep{laio2002,barducci2008} method was used to properly sample the equilibrium population of the various metal-ligand coordination species ($[ML_i]$). Besides, parallel-bias metadynamics (PBMETAD)\citep{Pfaendtner2015}, together with partition family setup\citep{Prakash2018}, was used to bias alternatively one of the metal-ligand/water coordination states. Up to 16 multiple-walkers (MW) \citep{Raiteri2006} were used to ensure convergence (Cd$^{2+}$ complex systems were simulated for 40 ns each MW with this setup, Ni$^{2+}$ complex system for 80 ns each MW). The deposited Gaussian bias potentials had initial height, width, and deposition stride equal to 1 kJ/mol, 0.1 nm, and 1 ps, respectively; the bias factor was set to 10. Metadynamics simulations were carried out using the open-source, community-developed PLUMED library (ver. 2.8)\citep{Plumed2019,tribello2014}. 
The free energy surfaces were computed using a reweighting procedure applied to the biased simulation trajectories. Upon convergence, the profiles obtained over the final 10 ns of biased simulations were averaged to determine the reported free energy values $\Delta G_{ij}$, with associated errors as standard deviations.\\ 
The minimum free energy pathways of complex formation were evaluated from the corresponding 2D free energy maps using the open-source software MEPSA (ver. 1.6)\citep{Alcalde2015}. The equilibrium population (i.e., concentration) of each complex species was obtained from:

\begin{equation}\label{eq:boltz_pop}
    \frac{n_i}{n_j}=exp^{-\Delta G_{ij}/k_BT}
\end{equation}

where $n_i$ is the equilibrium population associated with the $[ML_i]$ coordination species, $\Delta G_{ij}$ is the difference in free energy between the $i$-th and $j$-th coordination states, $k_B$ is the Boltzmann constant and $T$ the temperature. From equation \ref{eq:boltz_pop}, it can be shown that the association constants can be expressed in terms of $\Delta G_{ij}$ as:

\begin{equation}\label{eq:stab_constant}
    K_{i}=\frac{exp\left(-\Delta G_{ij}/k_BT\right)}{\left[L\right]}
\end{equation}

where the unknown free ligand concentration, $[L]$, can be determined from the mass conservation condition:

\begin{equation}\label{eq:free_ligand}
    [L] = [L_0] - \sum_{i=1}^{N_s} i[ML_i]
\end{equation}

where $[L_0]$ is the initial ligand concentration and $N_s$ the total number of possible coordination states. Errors relative to p$K_i$ were estimated using a Monte Carlo analysis \citep{Dowd1965, Bowser1998} by repeatedly sampling $\Delta G_{ij}$ from their probability distribution, assuming a normal distribution of these values with standard deviation equal to the error measured by profile average mentioned above. Finally, solving equation \ref{eq:stab_constant} to determine the statistical spread of $K_i$. Enthalpies ($\Delta H$) of complex formation were obtained from unbiased MD simulations of the bound and unbound state of each metal-ligand complex ($[ML_i]$) (i.e., $\Delta H^{bind}_i=H^{bound}_i-H^{unbound}_i$). Binding entropies ($\Delta S$) were obtained from the difference between $\Delta G$ and $\Delta H$.

\subsection*{Markov State Model}\label{sec:methods_MSM}

The characterization of the configurational space of the metal complexes described by the coordination number allowed the construction of a Markov State Model (MSM) to compute the rate constants between different coordination states. Initial structures were extracted every 0.5 step from the metal-water and metal-ligands coordination maps ($\approx 50$). From each of these structures 200 unbiased MD replicas were simulated randomly resampling the momenta. It is important to note that a similar approach using metadynamics to help build a reliable MSM has been used previously to explore the dynamics of the helical peptide Aib9 \citep{Biswas2018}. Each replica was 2 ns long and the coordination state for ion-ligand and ion-water was recorded every 100 fs. These data were processed using an in-house Python v.3.8 script with the help of the deeptime \citep{Hoffmann_2022} python library (see Code Availabilty for the software access). K-Means++ \citep{Arthur_2006} was used to find the lowest number of centers for the MSM that satisfied the implied timescale (see Fig. S\ref{fig:its}) and Chapman-Kolmogorov analysis (see Fig. S\ref{fig:ck-test}) \citep{prinz2011,Bowman_2016}. Successively PCCA+ \citep{Roblitz2013} was applied to reduce the MSM microstates (centers) to the actual experimental measurable metal complex states. The transition matrix T$_{ij}$ associated to the MSM was also reduced accordingly and mean first passage times (MFPT) were computed using transition path theory (TPT) \citep{Metzner2009} and the errors associated were computed using a full bayesian approach as described in \citep{noe2008, Bowman_2016}. \\
To compare with experimental rate constants computed MFPT needed to be transformed accordingly taking into account the free ligand concentration. 
Defining as in the results section, the kinetic rate $k_{\pm i}$ as the association ($+i$) or the dissociation rate ($-i$) for the formation or disruption of the i-th metal complex $ML_i$, thus from reaction law theory for the i-th complex state it can be written a kinetic equation of the form 

\begin{equation}
    \frac{d[ML_i]}{dt} = -k_{-i}[ML_i]+k_i[ML_{(i-1)}][L]-k_{i+1}[ML_{i}][L]+k_{-(i+1)}[ML_{i+1}]
\end{equation}

A similar kinetic equation can be written for the populations of the reduced MSM of the form

\begin{equation}
    \frac{dn_i}{dt} = -k_{-i}n_i+k_in_{(i-1)}-k_{i+1}n_{i}+k_{-(i+1)}n_{i+1}
\end{equation}

To have compatibility between the experimental and the computed kinetic models a one to one correspondence must be superimposed. Thus, the forward rate constants of the MSM must be multiplied by a factor $\gamma^{2}/n^{eq}_{lig}$ where $n^{eq}_{lig}$ is the number of unbounded ligand at equilibrium (eq. \ref{eq:free_ligand}) and $\gamma = N_{Av}V$, with $N_{Av}$ being the Avogadro's number and V the volume of the simulation box. Regarding the backward rate constants, the inverse of the MFPT calculated from the MSM must be multiplied just by a factor $\gamma$. This correction makes the model robust when dealing with different ion-ligand concentrations as can be seen in Supplementary Table \ref{tab:rates_Cden_diff_conc} where we are able to extract very similar kinetic rates for two different ligand concentrations.  

\subsection*{Mechanism analysis}\label{sec:method_mechanism}

To analyze the mechanism of metal complex formation, several unbiased trajectories were generated for 10 Cd$^{2+}$ and 30 en (10 ns each) and 10 Cd$^{2+}$ and 30 nme (5 ns each) systems, saving atomic positions every 0.1 ps. Distances between nitrogen atoms of the amines and Cd(II) ions were computed. To ignore spurious binding/unbinding events, a ligand was considered bound when its nitrogen atom(s) entered the ion first coordination shell (distance $<$ 3.4 \AA) and remained for at least 20 ps. \\
The binding events were classified by the change in Cd(II)-amine coordination (0-1, 1-2, 2-3, 3-4 bindings). The ion-ligand and ion-water coordination numbers were monitored for 20 ps before and after the binding events. The evolution of water/ligand coordination during the binding events was used to classify the ligand exchange mechanism (dissociative and associative). \\
Water-ligand exchange mechanism was also analyzed by monitoring the distances and angles between entering nitrogen, leaving water, and the ion during each binding event, following the work of Falkner et al. \citep{Falkner2021}.

\backmatter

\bmhead{Supplementary information}

\begin{itemize}
\item \textbf{Table S1}: Cd(II)-en stability constant at different concentrations 
\item \textbf{Table S2}: Cd(II)-en free energy difference at different concentrations 
\item \textbf{Table S3}: Computed stability constants
\item \textbf{Table S4}: Computed formation and dissociation rate constants 
\item \textbf{Table S5}: Simulation details 
\item \textbf{Table S6}: Cd(II)-nme stability constant at different polarizability 
\item \textbf{Table S7} Cd(II)-en stability constant from formation and dissociation kinetics 
\item \textbf{Table S8} Cd(II)-en formation and dissociation rate constants at different concentrations 
\item \textbf{Figure S1} Relative stability of Cd(II)-en complexes at different concentrations
\item \textbf{Figure S2} Water-ligand exchange mechanisms in Cd(en)$_2$ complex formation
\item \textbf{Figure S3} Thermodynamics of Ni(II)-nme complexes in aqueous solution
\item \textbf{Figure S4} Cd(II)-en implied timescales plot
\item \textbf{Figure S5} Cd(II)-en Chapman-Kolmogorov test

\end{itemize}

\bmhead{Acknowledgments}
We thank A. Sengupta of the NarangLab (UCLA) for assistance with the setup and implementation of the m12-6-4 for the systems studied. We gratefully acknowledge the computational resources and technical support of the Center for High Performance Computing (CHPC) at SNS and the CINECA under the ISCRA initiative (ISCRA-C projects: IONCHL23 and IONIDP24). 

\section*{Declarations}
\subsection*{Funding}
GB acknowledges financial support under the National Recovery and Resilience Plan (NRRP), Mission 4, Component 2, Investment 1.1, Call for tender No. 1409 published on 14.9.2022 by the Italian Ministry of University and Research (MUR), funded by the European Union – NextGenerationEU – Project AquaGreen – CUP E53D23015550001. KMM gratefully acknowledges financial support from the NIH (GM130641). 

\subsection*{Conflict of interest/Competing interests}
The authors declare no competing interests.
%\subsection*{Ethics approval}
%Not Applicable
%\subsection*{Consent to participate}
%Not Applicable
%\subsection*{Consent for publication}
%Not Applicable
\subsection*{Availability of data and materials}
Input files for reproducing MD simulations and enhanced sampling simulations described in the Methods section, together with a pipeline to reweight the results are available at the Zenodo repository at the link \url{https://zenodo.org/records/15230404}.
\subsection*{Code availability}
The code to compute equilibrium constants from free energy profiles issuing from enhanced sampling is available at \url{https://github.com/SNS-Brancato-Lab/metals-ligand-equilibrium.git}. \\
The code to perform the analysis of ligand-exchange mechanism is available at \url{https://github.com/SNS-Brancato-Lab/mechanism_ligand_exchange.git}. \\
The code that performs MSM calculation, evaluation and kinetic analysis is available at \url{https://github.com/SNS-Brancato-Lab/MSManalysis.git}
%\subsection*{Authors' contributions}
%Here contributions

\newpage

\setcounter{figure}{0} 
\begin{figure}[h!] 
\centering
\includegraphics[width=1.0\textwidth]{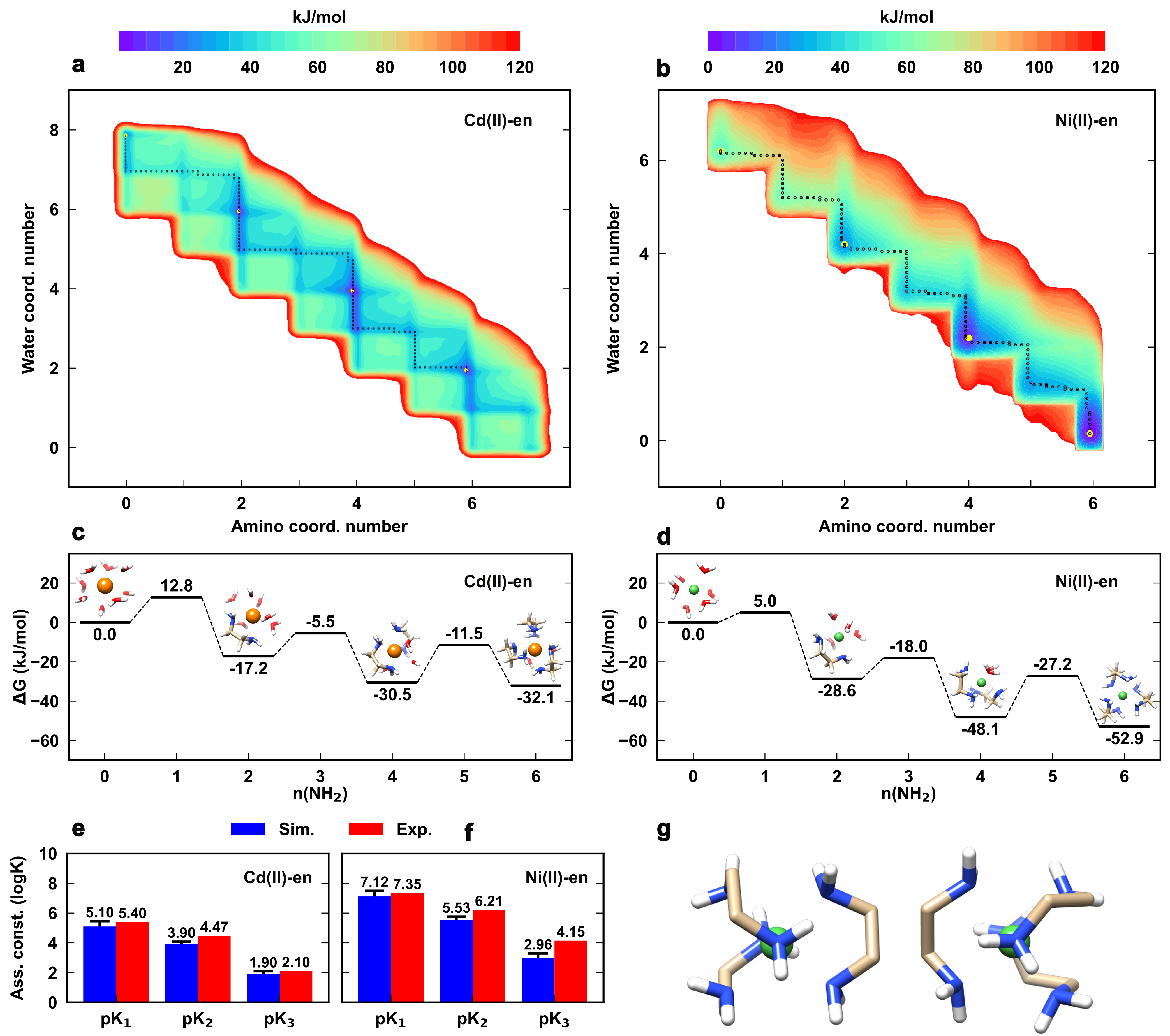}
\caption{ \textbf{Thermodynamics of Cd(II)-en and Ni(II)-en metal-ligand coordination in aqueous solution.} \textbf{a,b}, 2D free energy map of the Cd(II)-en (\textbf{a}) (Cd(II): 0.05M, en: 0.15M), and Ni(II)-en (\textbf{b}) (Ni(II): 0.05M, en: 0.15M) complex equilibrium in aqueous solution as a function of water and amino coordination number, showing all possible metal coordination states. Yellow points indicate the ML, ML$_2$, and ML$_3$ configurations. The dotted black lines are the minimum free energy pathways. Note that the profiles depend on the given concentration. \textbf{c,d}, Relative stability of the main Cd(II)-en (\textbf{c}) and Ni(II)-en (\textbf{d}) complex configurations with respect to the free metal ion. Estimated error is 2kJ/mol. \textbf{e,f}, Computed and experimental association constants (pK$_i$) of the ML, ML$_2$, and ML$_3$ complex species for the Cd(II)-en (\textbf{e}) and Ni(II)-en (\textbf{f}) systems. Estimated errors are reported in Supplementary Table \ref{tab:logKs_all}. \textbf{g}, Representative MD structures of the left-handed (left) and right-handed (right) enantiomers of the Ni(II)(en)$_3$ complex. 
\label{fig:fig1}}

\end{figure}

\begin{figure}[h!]%
\centering
\includegraphics[width=1.0\textwidth]{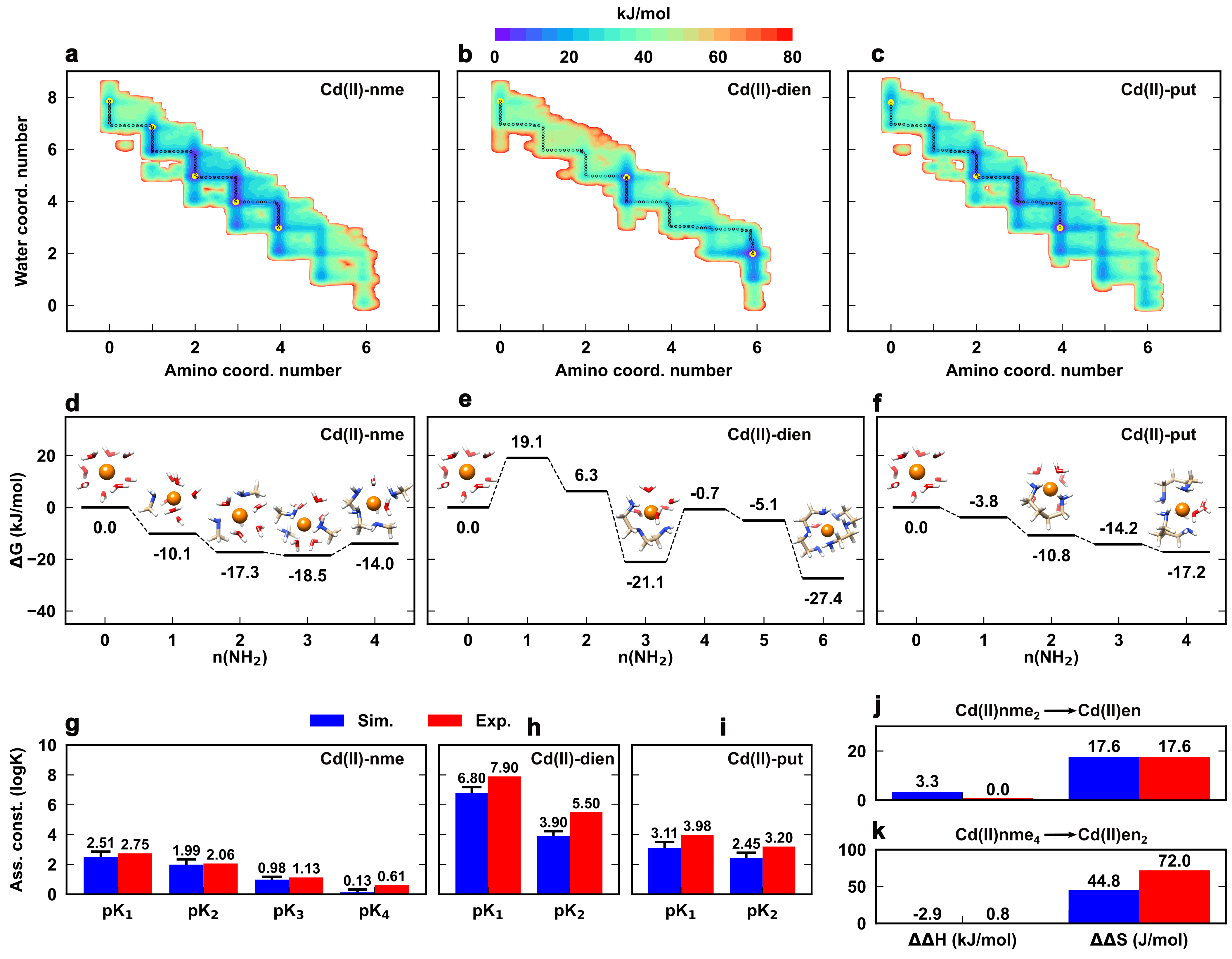}
\caption{\textbf{Thermodynamics of Cd(II) complexes with nme, dien, and put in aqueous solution.} \textbf{a-c}, 2D free energy map of the Cd(II)-nme (\textbf{a}) (Cd(II): 0.05M, nme: 0.30M), Cd(II)-dien (\textbf{b}) (Cd(II): 0.05M, dien: 0.10M), and Cd(II)-put (\textbf{c}) (Cd(II): 0.05M, put: 0.15) complex equilibrium in aqueous solution  as a function of water and amino coordination number, showing all possible metal coordination states. Yellow points indicate the ML, ML$_2$, ML$_3$, and ML$_4$ configurations. The dotted black lines are the minimum free energy pathway. Note that the profiles depend on the given concentration. \textbf{d-f}, Relative stability of the main Cd(II)-nme (\textbf{d}), Cd(II)-dien (\textbf{e}), and Cd(II)-put (\textbf{f}) complex configurations with respect to the free metal ion. Estimated error is 2kJ/mol. \textbf{g-i}, Computed and experimental association constants (pK$_i$) of the ML, ML$_2$, ML$_3$, and ML$_4$ complex species for the  \textbf{g} Cd(II)-nme, \textbf{h} Cd(II)-dien, and \textbf{i} Cd(II)-put systems. Estimated errors are reported in Supplementary Table \ref{tab:logKs_all}. \textbf{j,k}, Computed and experimental enthalpic and entropic energy differences between Cd(II)nme$_2$ and Cd(II)en formation (\textbf{j}), and between Cd(II)nme$_4$ and Cd(II)en$_2$ formation (\textbf{k}).\label{fig:fig2}}
\end{figure}

\begin{table}[h!]
\caption{Thermodynamic analysis of Cd(II) complexes with various ligands. $n$ indicates denticity. Values are in kJ/mol.}\label{tab:thermo-table}.
\begin{tabularx}{\textwidth}{*{7}{X}}
\toprule
Ligand$_{(i)}$ & Denticity & $\Delta$G  & $\Delta$H  & -T$\Delta$S  & $\Delta$H$/n$ & -T$\Delta$S$/n$ \\
\midrule
nme$_1$ & 1 & -14.2 & -23.0 & 8.8 & -23.0 & 8.8 \\
nme$_2$ & 2 & -26.8 & -46.9 & 20.1 & -23.5 & 10.1 \\
nme$_3$ & 3 & -32.6 & -63.2 & 30.5 & -21.1 & 10.2 \\
nme$_4$ & 4 & -33.1 & -81.6 & 48.5 & -20.4 & 12.1 \\
nme$_6$\footnotemark[1] & 6 & -15.9 & -98.7 & 82.8 & -16.5 & 13.8 \\
\midrule
en$_1$ & 2 & -29.3  & -43.5 & 14.2 & -21.8 & 7.1 \\
en$_2$ & 4 & -51.9 & -87.0 & 35.1 & -21.8 & 8.8 \\
en$_3$ & 6 & -62.7 & -119.2 & 56.4 & -19.9 & 9.4 \\
\midrule
dien$_1$ & 3 & -38.9 & -53.4 & 15.5 & -17.8 & 5.2 \\
dien$_2$ & 6 & -61.1 & -100.4 & 39.3 & -16.7 & 6.6 \\
\midrule
put$_1$ & 2 & -18.0 & -45.6 & 27.6  & -22.8 & 13.8 \\
put$_2$ & 4 & -31.8 & -89.1 & 57.3 & -22.3 & 14.3 \\
%\botrule
\end{tabularx}
\footnotetext[1]{Analysis based on configuration obtained from metadynamics simulations.}
\end{table}

\newpage

\begin{figure}[h!]%
\centering
\includegraphics[width=1.0\textwidth]{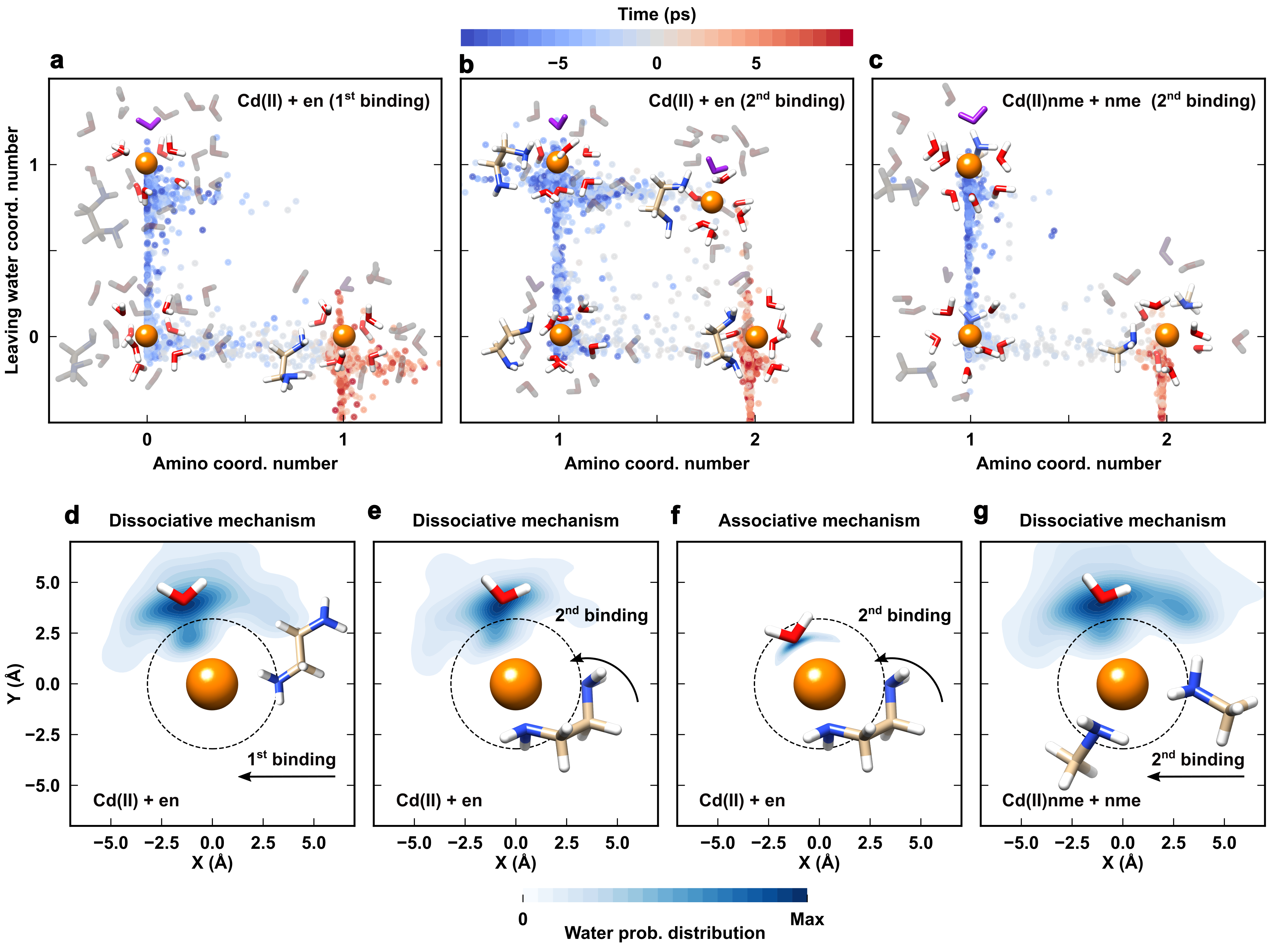}
\caption{\textbf{Water-ligand exchange mechanisms in Cd(II)-en and Cd(II)-nme complexes.} \textbf{a-c}, Evolution of amino and leaving water coordination number in the first (\textbf{a}) and second (\textbf{b}) binding event of Cd(II)en formation, and in the second binding event (\textbf{c}) of Cd(II)nme$_2$ formation. The leaving water is represented in purple.
\textbf{d-g}, Probability distribution of the leaving water in the time interval from 5 ps before to 5 ps after the first dissociative binding (\textbf{d}), the second dissociative (\textbf{e}) and second associative (\textbf{f}) binding events of Cd(II)en formation, and the second dissociative binding event of Cd(II)nme$_2$ formation (\textbf{g}). The black dotted circle represents the Cd(II) first solvation shell.\label{fig:fig3}}
\end{figure}

\begin{figure}[b!]%
\centering
\includegraphics[width=1.0\textwidth]{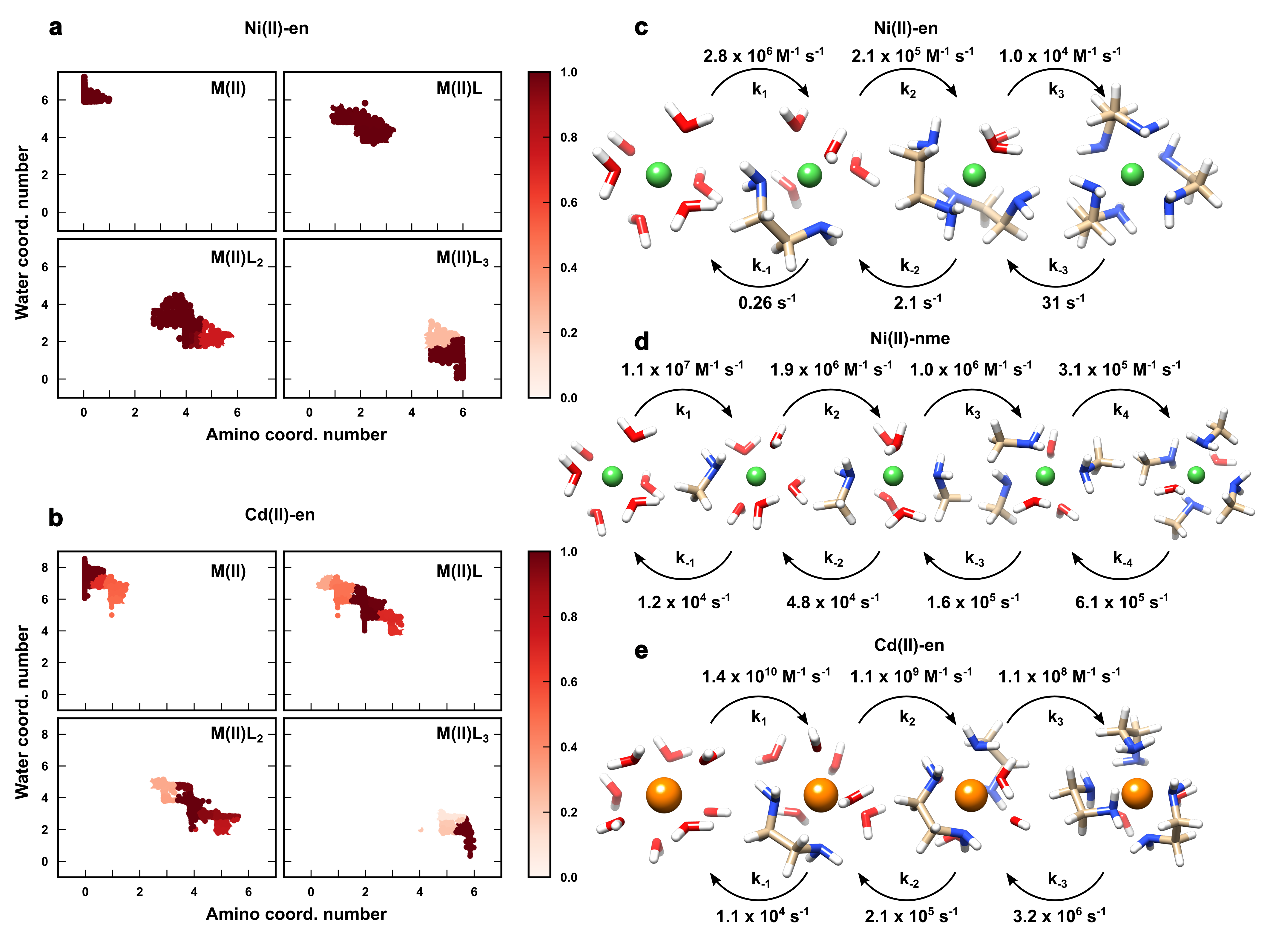}
\caption{\textbf{Kinetic analysis of ligand binding and unbinding in Cd(II) and Ni(II) complexes.}\textbf{a,b}, Macrostate assignment probablity of the short unbiased MD trajectories following PCCA+ analysis for Ni(II)-en (\textbf{a}) and Cd(II)-en (\textbf{b}). The four macrostates identified by the algorithm accurately correspond to distinct coordination numbers (ML$_i$) of the metal ion with the ethylenediamine.
\textbf{c-e}, Computed formation (k$_i$) and dissociation (k$_{-i}$) rate constants for Ni(II)-en (\textbf{c}), Ni(II)-nme (\textbf{d}), and Cd(II)-en (\textbf{e}). Estimated errors are reported in Supplementary Table \ref{tab:rates_all}. 2D free energy map, relative stability of the main complexes and association constant (pK$_i$) of Ni(II)-nme system are shown in Supplementary Fig. \ref{fig:ni-nme}. \label{fig:fig4}}
\end{figure}

\clearpage
\newpage

\begin{figure}[b!]%
\centering
\includegraphics[width=1.0\textwidth]{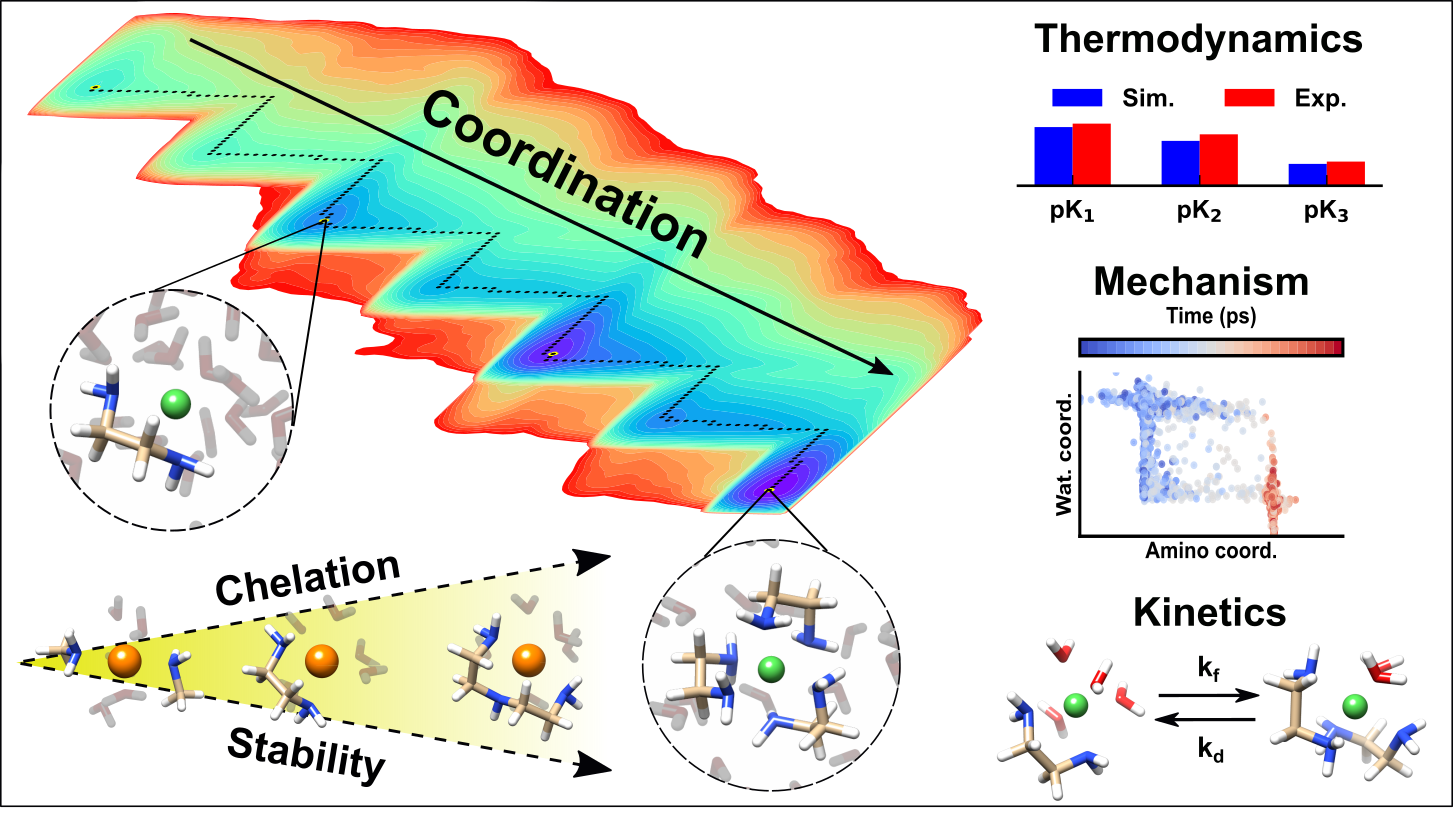}
\caption{Graphical abstract}
\end{figure}

\clearpage
\newpage

\bibliography{bibliography}% common bib file

\makeatletter\@input{xx2.tex}\makeatother
\end{document}

\title[Article Title]{\qquad \quad Supplementary Information for:\\Simulating metal complex formation and dynamics in aqueous solutions: Insights into stability, mechanism, and rates of ligand exchange}


\author[1,2,3]{\fnm{Luca} \sur{Sagresti}}%\email{iauthor@gmail.com}

\author[1,2]{\fnm{Luca} \sur{Benedetti}}
%Uncomment line below for equal contribution

\author[4,5]{\fnm{Kenneth M.} \sur{Merz Jr.}}
%Uncomment line below for equal contribution

\author*[1,2,3]{\fnm{Giuseppe} \sur{Brancato}}\email{giuseppe.brancato@sns.it}
%Uncomment line below for equal contribution

\affil[1]{\orgname{Scuola Normale Superiore}, \orgaddress{\street{Piazza Dei Cavalieri 7}, \city{Pisa}, \postcode{I-56126}, \country{Italy}}}
\affil[2]{\orgname{Istituto Nazionale di Fisica Nucleare (INFN)}, \orgaddress{\street{Largo Pontecorvo 3}, \city{Pisa}, \postcode{I-56127}, \country{Italy}}}

\affil[3]{\orgname{Consorzio Interuniversitario per Lo Sviluppo Dei Sistemi a Grande Interfase (CSGI)}, \orgaddress{\street{Via Della Lastruccia 3}, \city{Sesto Fiorentino (Fi)}, \postcode{I-50019}, \country{Italy}}}

\affil[4]{\orgdiv{Department of Chemistry}, \orgname{Michigan State University}, \orgaddress{\street{Street}, \city{East Lansing}, \postcode{48824}, \state{Michigan}, \country{United States}}}

\affil[5]{\orgdiv{Department of Biochemistry and Molecular Biology}, \orgname{Michigan State University}, \orgaddress{\street{Street}, \city{East Lansing}, \postcode{48824}, \state{Michigan}, \country{United States}}}

\maketitle

\newpage

\tableofcontents
\listoftables
\listoffigures

\newpage

%% Tab 1
\begin{table}[h]
\caption[Cd(II)-en stability constant at different concentrations]{Stability constants ($pK_{i}$) between different metal ligand coordination states as computed following  section 4.2 at different cadmium and ethylenediamine concentrations}\label{tab:pKs_diffConc}
\begin{tabularx}{\textwidth}{*{4}{X}}
\toprule
Cd(II)-en & p$K_1$ & p$K_2$ & p$K_3$ \\
\midrule
exp\footnotemark[1]  &  5.4  & 4.47 & 2.1 \\ 
1-3   &  5.2 $\pm$ 0.4 & 4.0 $\pm$ 0.2 & 1.8 $\pm$ 0.3\\
2-6   &  5.1 $\pm$ 0.4 & 4.1 $\pm$ 0.2 & 1.8 $\pm$ 0.3\\
10-45 &  5.1 $\pm$ 0.4 & 3.8 $\pm$ 0.2 & 1.7 $\pm$ 0.2\\
10-30 &  5.1 $\pm$ 0.3 & 3.9 $\pm$ 0.2 & 1.9 $\pm$ 0.2\\
\botrule
\end{tabularx}
\footnotetext[1]{Ref. \citep{Paoletti1984}}
\end{table}

\newpage

%% Tab 2
\begin{table}[h]
\caption[Cd(II)-en free energy difference at different concentrations]{Energy difference from free energy profile ($\Delta G_{ij}$) between different metal ligand coordination states at different cadmium and ethylenediamine concentrations. Between parenthesis the value that it would be expected if taking experimental stability constants as starting point to reverse equation 3 in section 4.2}\label{tab:Fs_diffConc}
\begin{tabularx}{\textwidth}{*{4}{X}}      %{@{}llll@{}}
\toprule
Cd(II)-en & $\Delta G_{01}$ (kJ/mol) & $\Delta G_{12}$ (kJ/mol) & $\Delta G_{23}$ {kJ/mol} \\
\midrule
1-3   &  18.5 (19.5)  & 12.8 (14.2)  & -0.3 (1.3) \\
2-6   &  18.4 (19.4)  & 12.7 (14.1)  & -0.3 (1.2) \\
10-45 &  23.7 (24.9)  & 15.6 (19.5)  & 3.6 (6.7) \\
10-30 &  17.2 (20.4)  & 13.3 (15.1)  & 1.6 (2.4) \\
10-20 &  15.4 (12.9)  & 3.8 (7.0)    & -5.2 (-7.4) \\
10-10 &  5.0 (2.7)    & -5.2 (-2.7)  & -26.1 (-15.5) \\
\textbf{MAE} & 1.9    & 2.4          & 3.9 \\
\botrule
\end{tabularx}
\end{table}

\newpage

%% Tab 3
\begin{table}[h]
\caption[Computed stability constants]{Stability constants (p$K_{i}$) between different metal ligand coordination states for each system under investigation. Experimental values, where available, are between parenthesis.}\label{tab:logKs_all}
\begin{tabularx}{\textwidth}{*{5}{X}}
\toprule
M-L & p$K_1$ & p$K_2$ & p$K_3$ & p$K_4$\\
\midrule 
Cd(II)-en   &  5.10 $\pm$ 0.35 & 3.90 $\pm$ 0.18  & 1.90 $\pm$ 0.19 & \\
 & (5.40)\footnotemark[1] & (4.47)\footnotemark[1] & (2.10)\footnotemark[1] & \\
\\
Cd(II)-nme & 2.51 $\pm$ 0.35 & 1.99 $\pm$ 0.35 & 0.98 $\pm$ 0.19 & 0.13 $\pm$ 0.19\\
 & (2.75)\footnotemark[2] & (2.06)\footnotemark[2] & (1.13)\footnotemark[2] & (0.61)\footnotemark[2] \\
\\
Cd(II)-dien & 6.80 $\pm$ 0.39 & 3.90 $\pm$ 0.33 & & \\
 & (7.90)\footnotemark[3] & (5.50)\footnotemark[3] & & \\
\\
Cd(II)-put & 3.11 $\pm$ 0.40 & 2.45 $\pm$ 0.34 & & \\
 & (3.98)\footnotemark[3] & (3.20)\footnotemark[3] & & \\
\\
Ni(II)-en & 7.12 $\pm$ 0.38 & 5.53 $\pm$ 0.23 & 2.96 $\pm$ 0.33 & \\
 & (7.35)\footnotemark[4] & (6.21)\footnotemark[4] & (4.15)\footnotemark[4] & \\
\\
Ni(II)-nme & 2.16 $\pm$ 0.37 & 1.48$\pm$ 0.19 & 0.55 $\pm$ 0.19 & -0.86 $\pm$ 0.53 \\
 & (2.23)\footnotemark[5] & & & \\
\botrule
\end{tabularx}
\footnotetext[1]{Ref. \cite{Paoletti1984}}
\footnotetext[2]{Ref. \cite{Spike1953}}
\footnotetext[3]{Ref. \cite{Pettit2006}}
\footnotetext[4]{Ref. \cite{Taylor1974}}
\footnotetext[5]{Ref. \cite{Rorabacher1971}}
\end{table}

\newpage

%% Tab 4
\begin{table}[h]
\caption[Computed formation and dissociation rate constants]{Formation and dissociation rate constants ($k_{i}$) between different metal ligand coordination states in systems under investigation.}\label{tab:rates_all}
\begin{tabularx}{\textwidth}{*{5}{X}}
\toprule
ML & $k_1$ ($M^{-1}s^{-1}$) & $k_2$ ($M^{-1}s^{-1}$) & $k_3$ ($M^{-1}s^{-1}$) & $k_4$ ($M^{-1}s^{-1}$)\\
\midrule
Cd(II)-en & 1.4 $\pm$ 0.2 $\cdot$ 10$^{10}$ & 1.1 $\pm$ 0.1 $\cdot$ 10$^{9}$ & 1.1 $\pm$ 0.1 $\cdot$ 10$^{8}$ & \\
&  &  &  & \\
Cd(II)-nme & 1.2 $\pm$ 0.1 $\cdot$ 10$^{10}$ & 1.3 $\pm$ 0.1 $\cdot$ 10$^{10}$ & 6.5 $\pm$ 0.1 $\cdot$ 10$^{9}$ & 1.4 $\pm$ 0.1 $\cdot $10$^{9}$\\
&  &  &  & \\
Ni(II)-en & 2.8 $\pm$ 0.3 $\cdot$ 10$^{6}$  & 2.1 $\pm$ 0.3 $\cdot$ 10$^{5}$  & 1.0 $\pm$ 0.1 $\cdot$ 10$^{4}$  & \\
&  &  &  & \\
Ni(II)-nme & 1.1 $\pm$ 0.1 $\cdot$ 10$^{7}$  & 1.9 $\pm$ 0.1 $\cdot$ 10$^{6}$ & 1.0 $\pm$ 0.1 $\cdot$ 10$^{6}$ & 3.1 $\pm$ 0.3 $\cdot$ 10$^{5}$ \\
&  &  &  & \\
\midrule
& $k_{-1}$ ($s^{-1}$) & $k_{-2}$ ($s^{-1}$) & $k_{-3}$ ($s^{-1}$) & $k_{-4}$ ($s^{-1}$)\\
\midrule
Cd(II)-en& 1.1 $\pm$ 0.1 $\cdot$ 10$^{4}$ & 2.1 $\pm$ 0.1 $\cdot$ 10$^{5}$  & 3.2 $\pm$ 0.1 $\cdot$ 10$^{6}$  &\\
&  &  &  & \\
Cd(II)-nme& 2.1 $\pm$ 0.1 $\cdot$ 10$^{6}$  & 8.4 $\pm$ 0.1 $\cdot$ 10$^{7}$  & 8.4 $\pm$ 0.1 $\cdot$ 10$^{8}$  & 3.7 $\pm$ 0.1 $\cdot$ 10$^{9}$\\
&  &  &  & \\
Ni(II)-en& 0.26 $\pm$ 0.1 & 2.1 $\pm$ 0.3 & 31 $\pm$ 4& \\
&  &  &  & \\
Ni(II)-nme& 1.2 $\pm$ 0.1 $\cdot$ 10$^{4}$ & 4.8 $\pm$ 0.6 $\cdot$ 10$^{4}$ & 1.6 $\pm$ 0.1 $\cdot$ 10$^{5}$ & 6.1 $\pm$ 0.4 $\cdot$ 10$^{5}$\\
&  &  &  & \\
\botrule
\end{tabularx}
\end{table}

\newpage

%% Tab 5
\begin{table}[h]
\caption[Simulation details]{Physical and atomistic details of all the systems simulated.}\label{tab:sim-recap}
\begin{tabularx}{\textwidth}{*{5}{X}}
\toprule
 System & \#Metals & \#Ligands & Volume(nm$^3$) & Ligand Conc.(M) \\
\midrule
Cd-en  &  1 & 3  & 64 & 0.08 \\ 
Cd-en  &  2 & 6  & 128 & 0.08 \\
Cd-en  &  10 & 10 & 330 & 0.05  \\
Cd-en  &  10 & 20 & 330  & 0.10 \\
Cd-en  &  10 & 30 & 330  & 0.15 \\
Cd-en  &  10 & 45 & 330  & 0.22 \\
Cd-nme &   1 & 6  & 64  & 0.16 \\
Cd-nme &  10 & 60 & 330 & 0.30 \\
Cd-dien &  10 & 20 & 330  & 0.10 \\
Cd-put &  10 & 30 & 330  & 0.15 \\
Ni-en  &  1 & 3 & 64 & 0.08 \\
Ni-nme &  1 & 6 & 64 & 0.16 \\

\botrule
\end{tabularx}
\end{table}

\newpage

%% Tab 6
\begin{table}[h]
\caption[Cd(II)-nme stability constant at different polarizability]{Stability constants ($pK_{i}$) between different metal ligand coordination states as computed following section 4.2 at cadmium and methylamine concentrations (0.30 M) using different polarizability values for the 12-6-4 LJ model between Cd(II) and nme. }\label{tab:pKs_diff-nme}
\begin{tabularx}{\textwidth}{*{5}{X}}
\toprule
 & $pK_1$ & $pK_2$ & $pK_3$ & $pK_4$ \\
\midrule
exp\footnotemark[1]    &  2.75  & 2.06  & 1.13 & 0.61 \\ 
$\alpha_{Cd-nme}$=3.40\footnotemark[2]  &  2.51 & 1.99  & 0.98 & 0.13 \\
$\alpha_{Cd-nme}$=3.15\footnotemark[3]  & 2.10 & 1.21 & 0.22  & -0.81 \\
\botrule
\end{tabularx}
\footnotetext[1]{Ref. \citep{Spike1953}}
\footnotetext[2]{Polarizability parameter used in the work.}
\footnotetext[3]{Polarizability parameter from ref. \citep{Sengupta2018}.}
\end{table}

\newpage

\begin{table}[h]
\caption[Cd(II)-en stability constant from formation and dissociation kinetics]{Stability constants ($pK_{i}$) between different metal ligand coordination states for two different Cd-en concentrations (0.08 M and 0.15 M) as computed from the ratio of the formation and dissociation rate constants of table \ref{tab:rates_Cden_diff_conc}. }\label{tab:pKs_Cden_from_rates}%
\begin{tabularx}{\textwidth}{*{4}{X}}
\toprule
Ligand conc. & $pK_1$ & $pK_2$ & $pK_3$ \\
\midrule
0.08 M &  6.1 $\pm$ 0.8 & 4.2 $\pm$ 0.1 & 2.09 $\pm$ 0.05  \\
0.15 M &  6.1 $\pm$ 0.5 & 3.8 $\pm$ 0.1 & 1.55 $\pm$ 0.01  \\
\botrule
\end{tabularx}
\end{table}

\newpage

\begin{table}[h]
\caption[Cd(II)-en formation and dissociation rate constants at different concentrations]{Formation and dissociation rates ($k_{i}$) between different metal ligand coordination states between 2 different concentrations of Cd(II)-en: 0.08M and 0.15.}\label{tab:rates_Cden_diff_conc}
\begin{tabularx}{\textwidth}{*{4}{X}}
\toprule
Ligand conc. & $k_1$ ($M^{-1}s^{-1}$) & $k_2$ ($M^{-1}s^{-1}$) & $k_3$ ($M^{-1}s^{-1}$) \\
\midrule
0.08 M &  4.1 $\pm$ 1.1 $\cdot$ 10$^{10}$  & 6.6 $\pm$ 0.1 $\cdot$ 10$^9$  &  3.89 $\pm$ 0.08 $\cdot$ 10$^8$ \\
0.15 M & 1.4 $\pm$ 0.2 $\cdot$ 10$^{10}$ & 1.14 $\pm$ 0.02 $\cdot$ 10$^9$ & 1.09 $\pm$ 0.02 $\cdot$ 10$^8$  \\
\midrule
& $k_{-3}$ ($s^{-1}$) & $k_{-2}$ ($s^{-1}$) & $k_{-1}$ ($s^{-1}$) \\
\midrule
0.08 M & 3.17 $\pm$ 0.07 $\cdot$ 10$^6$ & 4.2 $\pm$ 0.7 $\cdot$ 10$^5$ & 3.5 $\pm$ 0.6 $\cdot$ 10$^4$ \\
0.15 M & 3.16 $\pm$ 0.06 $\cdot$ 10$^6$ & 2.05 $\pm$ 0.04 $\cdot$ 10$^5$ & 1.1 $\pm$ 0.1 $\cdot$ 10$^4$ \\
\botrule
\end{tabularx}
\end{table}

\newpage

%% Fig 1
\begin{figure}[h]%
\centering
\includegraphics[width=0.9\textwidth]{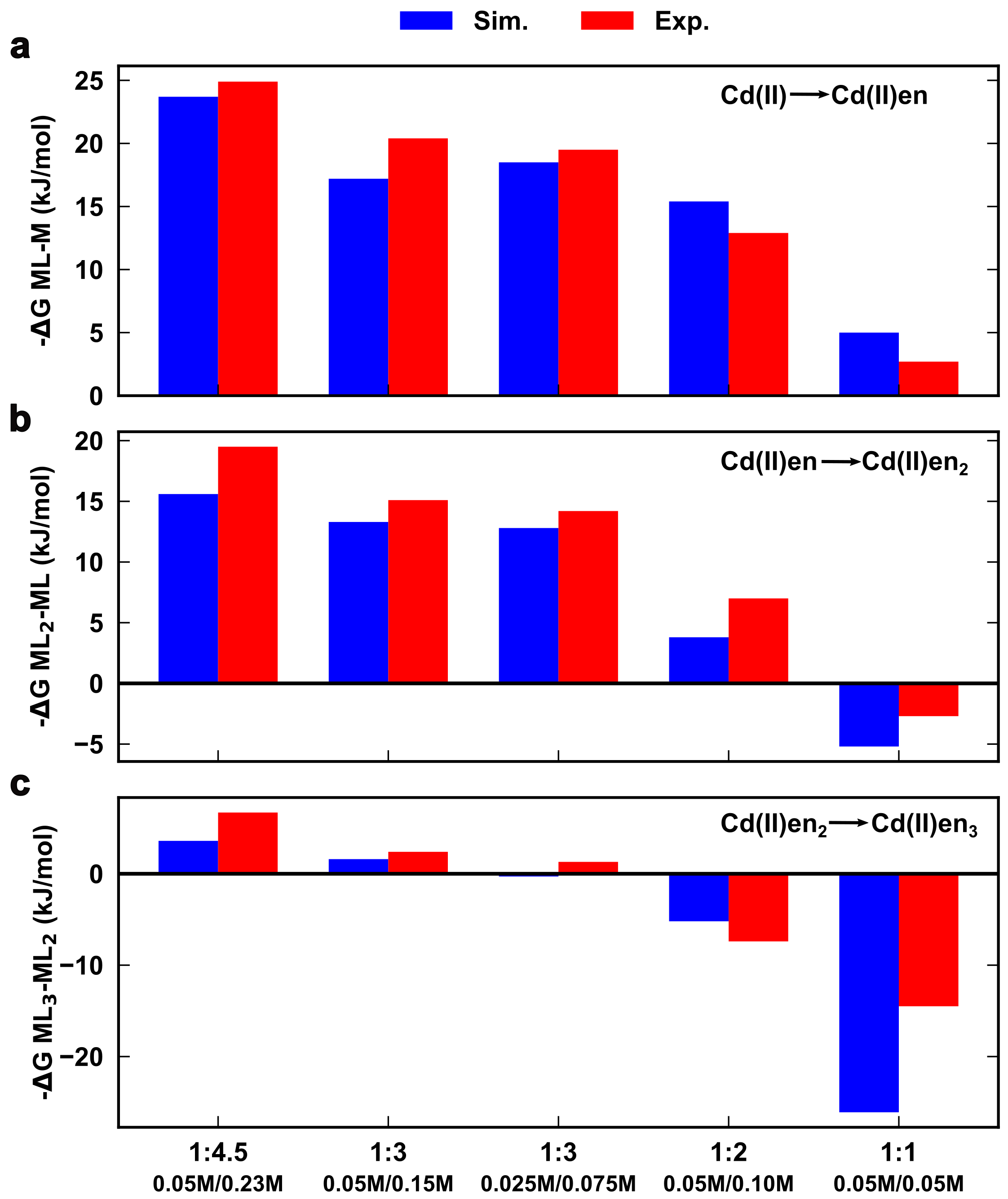}
\caption[Relative stability of Cd(II)-en complexes at different concentrations]{\textbf{a-c}, Relative stability between fully solvated Cd(II) and Cd(en) complex, $\Delta G^{ML-M}$ (\textbf{a}), between Cd(en) and Cd(en)$_2$, $\Delta G^{ML_2-ML}$ (\textbf{b}), and between Cd(en)$_2$ and Cd(en)$_3$, $\Delta G^{ML_3-ML_2}$ (\textbf{c}). Results are shown at different concentrations. In blue the computed results and in red the experimental measured ones.}\label{fig:multi-Cd-en}
\end{figure}

\newpage

%% Fig 2
\begin{figure}[h]%
    \centering
    \includegraphics[width=1\linewidth]{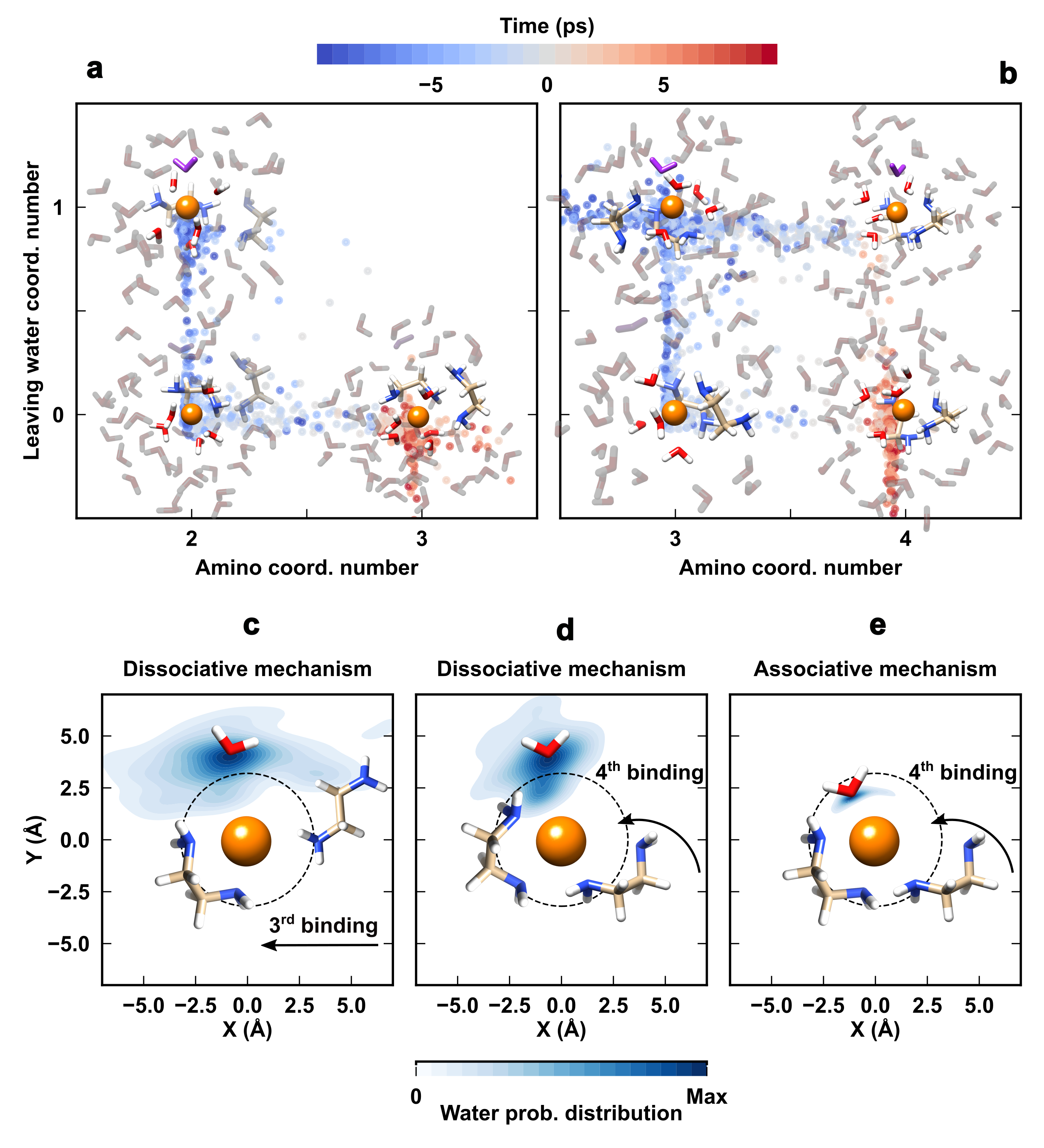}
    \caption[Water-ligand exchange mechanisms in Cd(en)$_2$ complex formation]{\textbf{a,b}, Evolution of amino and leaving water coordination number in the third (\textbf{a}) and fourth (\textbf{b}) binding event of Cd(II)en$_2$ formation. The leaving water is represented in purple.
    \textbf{c-e}, Probability distribution of the leaving water in the time interval from 5 ps before to 5 ps after the third dissociative binding (\textbf{c}), fourth dissociative (\textbf{d}) and fourth associative (\textbf{e}) binding events of Cd(II)en$_2$ formation. The black dotted circles represent the Cd(II) first solvation shell.}
    \label{fig:mechanism_en}
\end{figure}

\newpage

% Fig 3
\begin{figure}[h]%
    \centering
    \includegraphics[width=1\linewidth]{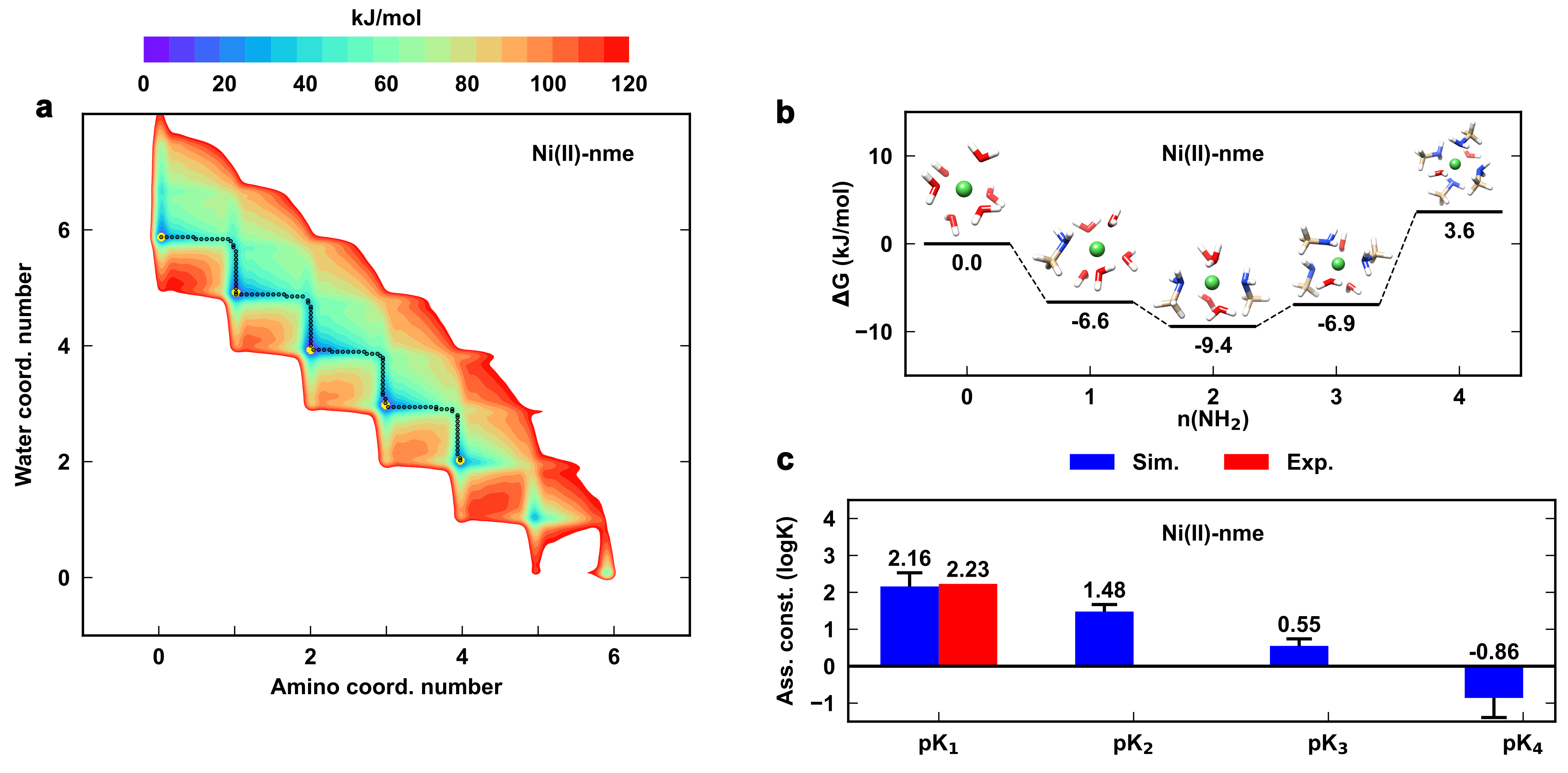}
    \caption[Thermodynamics of Ni(II)-nme complexes in aqueous solution]{\textbf{a}, 2D free energy map of the Ni(II)-nme complex equilibrium in aqueous solution (Ni(II): 0.10M, en: 0.60M), as a function of water and amino coordination number, showing all possible metal coordination states. Yellow points indicate the ML, ML$_2$, ML$_3$ and ML$_4$ configurations. The dotted black line is the minimum free energy pathway. Note that the profile depends on the given concentration. \textbf{b}, Relative stability of the main Ni(II)-nme complex configurations with respect to the free metal ion. Estimated errors is 2 kJ/mol. \textbf{c}, Computed and experimental association constants (pK$_i$) of the ML, ML$_2$, ML$_3$ and ML$_4$ complex species for the Ni(II)-nme system. Estimated errors are reported in Supplementary Table \ref{tab:logKs_all}. }\label{fig:ni-nme}
\end{figure}

\newpage

% METHODS KINETICS
% Fig 4
\begin{figure}[h]%
\centering
\includegraphics[width=1.0\textwidth]{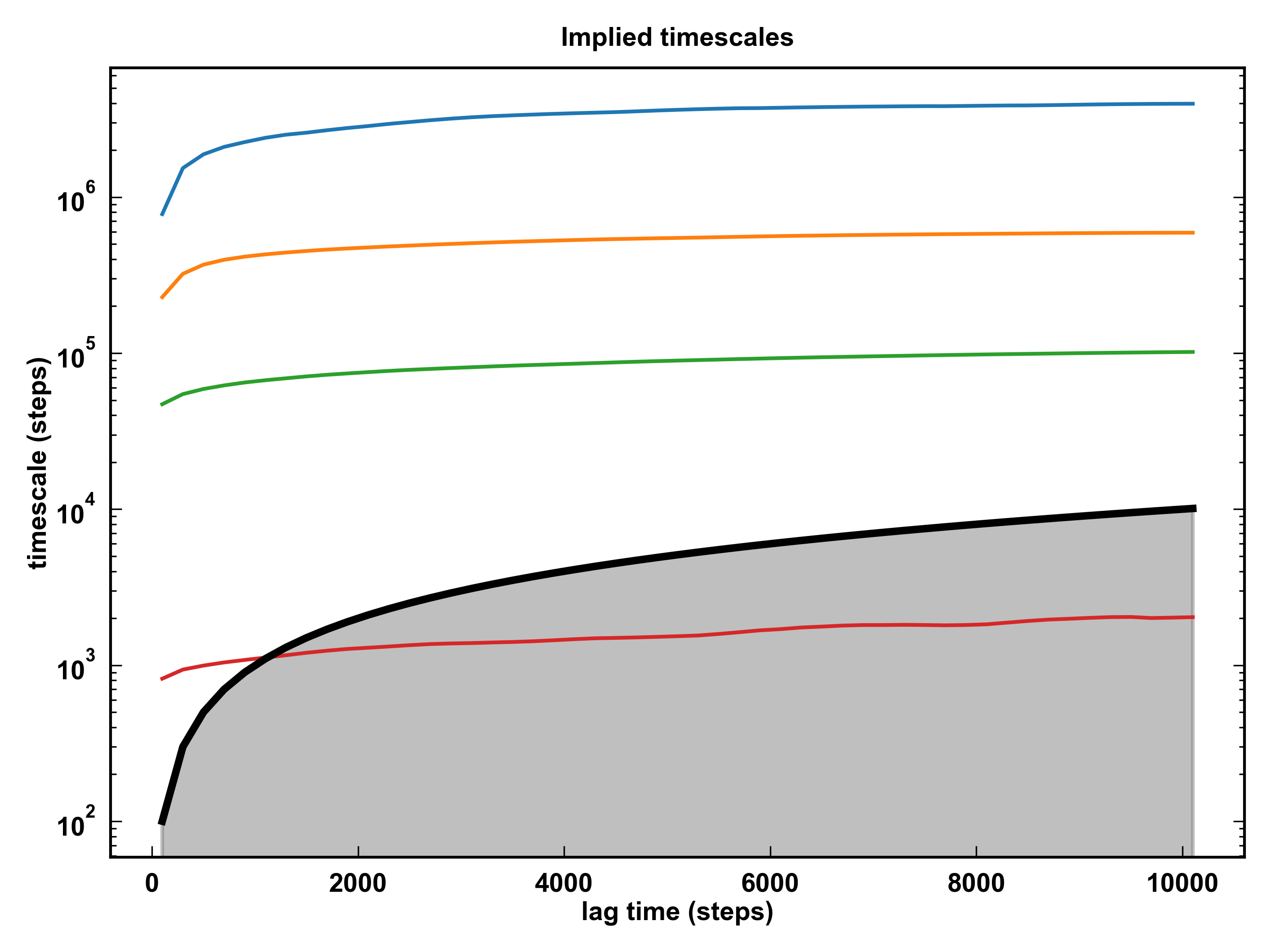}
\caption[Cd(II)-en implied timescale plot]{Implied timescales plot for the 4 slowest eigenvalues associated to the Markov state model of the Cd(II)-en system. After 600 steps (60 ps) the three slowest implied timescales are nicely approximated even for longer lagtimes.}\label{fig:its}
\end{figure}

\newpage

% METHODS KINETICS
% Fig 5
\begin{figure}[h]%
\centering
\includegraphics[width=1.2\textwidth]{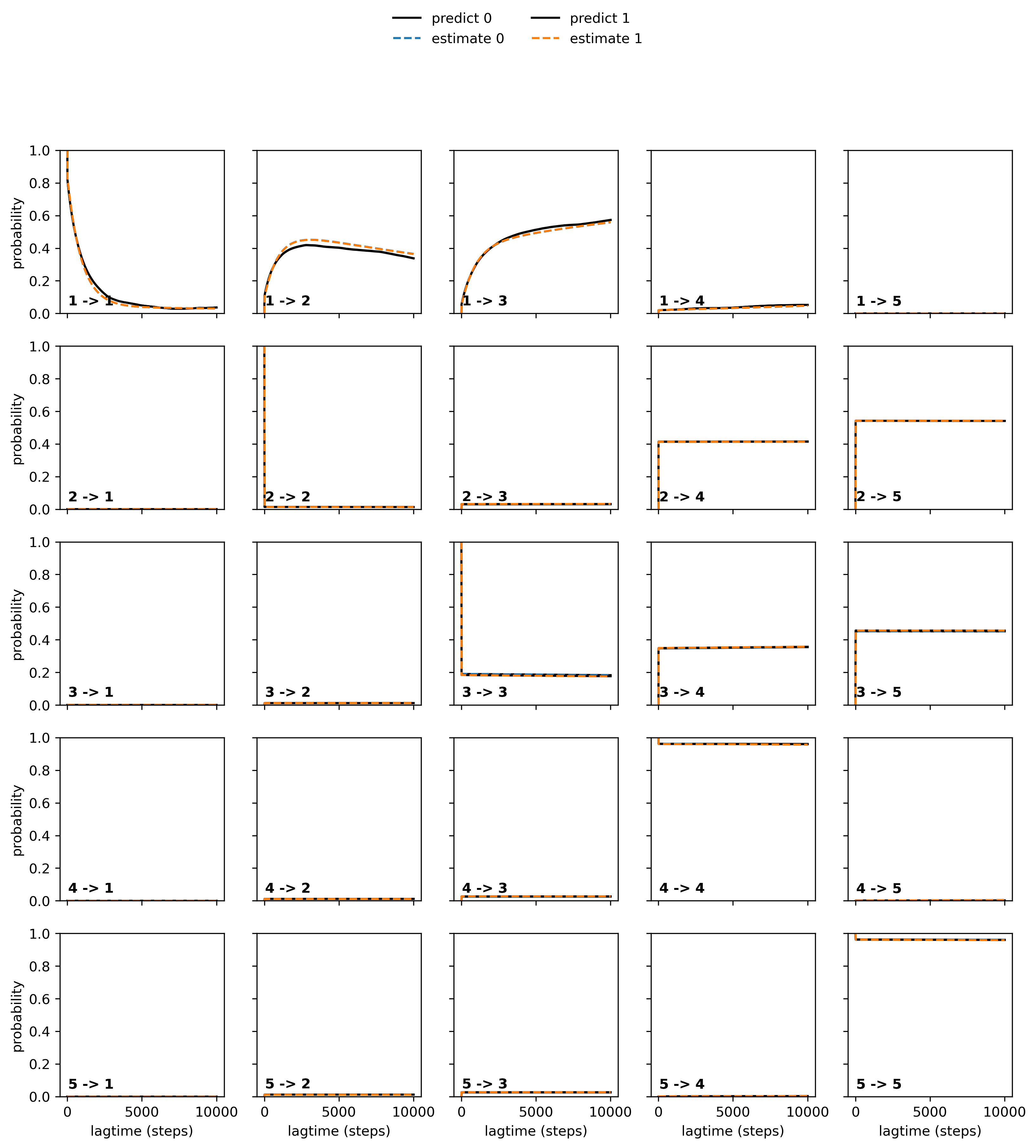}
\caption[Cd(II)-en Chapman-Kolmogorov test]{Chapman-Kolmogorov test of the Cd(II)-en Markov state model at 60 ps lagtime. The dynamics of the 5 metastable states at longer lagtimes are well reproduced for the lagtime chosen (60 ps). }\label{fig:ck-test}
\end{figure}

\clearpage
\bibliography{bibliography}% common bib file